\pdfoutput=1
\documentclass[10pt,conference]{IEEEtran}
\IEEEoverridecommandlockouts

\usepackage{cite}
\usepackage{amsmath,amssymb,amsfonts}
\usepackage{graphicx}
\usepackage[table]{xcolor}
\usepackage{booktabs}
\usepackage{multirow}
\usepackage{makecell}
\usepackage{threeparttable}
\usepackage{enumitem}
\usepackage[hyphens]{url}
\usepackage{hyperref}
\usepackage{xspace}
\usepackage{listings}
\usepackage{tcolorbox}

\definecolor{codebg}{RGB}{242,242,242}
\definecolor{codeline}{RGB}{170,170,170}
\definecolor{codecomment}{RGB}{0,128,0}
\definecolor{codekeyword}{RGB}{0,128,128}
\definecolor{codefunc}{RGB}{52,148,186}
\definecolor{codeop}{RGB}{0,128,128}

\lstdefinestyle{qiskitpython}{
    language=Python,
    basicstyle=\linespread{1.08}\ttfamily\footnotesize,
    backgroundcolor=\color{codebg},
    numbers=left,
    numberstyle=\tiny\color{codeline},
    numbersep=6pt,
    xleftmargin=1.2em,
    frame=leftline,
    rulecolor=\color{codeline},
    framesep=4pt,
    columns=fullflexible,
    keepspaces=true,
    showstringspaces=false,
    breaklines=true,
    tabsize=4,
    aboveskip=2pt,
    belowskip=2pt,
    keywordstyle=\color{codekeyword},
    commentstyle=\color{codecomment},
    stringstyle=\color{black},
    emph={execute,transpile,run},
    emphstyle=\color{codefunc},
    morekeywords={from,import},
}

\newcommand{\figref}[1]{Figure~\ref{#1}}
\newcommand{\tabref}[1]{Table~\ref{#1}}

\definecolor{commentgreen}{rgb}{0.0,0.38,0.16}

\newtcolorbox{summaryboxenv}{%
  colback=gray!10,
  colframe=black,
  boxrule=0.4pt,
  arc=0pt,
  outer arc=0pt,
  left=4pt, right=4pt, top=3.5pt, bottom=3.5pt,
  boxsep=0pt,
  before skip=5pt,
  after skip=5pt,
}
\newcommand{\summaryblock}[2]{%
\begin{summaryboxenv}
\noindent \textbf{#1:} #2
\end{summaryboxenv}}

\begin{document}

\title{PreMaQ: Predicting Maintainability-Related Quality of LLM-Generated Code Before Generation}

\author{
\IEEEauthorblockN{
Taiki Wakamatsu\IEEEauthorrefmark{1},
Yuta Ishimoto\IEEEauthorrefmark{2},
Masanari Kondo\IEEEauthorrefmark{1},
Yasutaka Kamei\IEEEauthorrefmark{1}
}
\IEEEauthorblockA{
\IEEEauthorrefmark{1}Kyushu University, Fukuoka, Japan \quad
\IEEEauthorrefmark{2}The University of Osaka, Osaka, Japan \\
t.wakamatsu@posl.ait.kyushu-u.ac.jp,\quad
ishimoto@ist.osaka-u.ac.jp,\quad
\{kondo, kamei\}@ait.kyushu-u.ac.jp
}
}

\maketitle

\begin{abstract}
As large language models (LLMs) become increasingly capable of code generation, adopting generated code in software development requires assessing not only its functional correctness but also its maintainability-related quality.
If such quality could be estimated before generation, developers could avoid the cost of generating, reviewing, and discarding low-quality code.
Although prior work has shown that the functional correctness of the LLM-generated code can be predicted in advance, it remains unclear whether maintainability-related quality is similarly predictable.
We introduce Pre-Generation Maintainability-Related Quality Prediction (PreMaQ), which predicts the Code Smell Score (CSS) and Maintainability Index (MI) of generated code from the internal representations of LLMs before generation.
Our evaluation covers four open-weight LLMs and four Python code generation benchmarks, comprising 2,695 tasks in total.
Our results show that predicted CSS and MI consistently correlate with their observed values across all 16 model–benchmark combinations, achieving mean Spearman rank correlations of 0.57 and 0.65, respectively.
When used for model selection, PreMaQ achieves 59.5\% of the maintainability-related quality improvement attainable by an ideal maintainability-based selector over random selection on tasks for which multiple models generate functionally correct code.
Combining predictions from PreMaQ and prompt embeddings increases this proportion to 60.7\%, indicating that the two signals are complementary.
These findings suggest that PreMaQ can be used to predict and improve the maintainability-related quality of LLM-generated code.

\end{abstract}


\section{Introduction}\label{sec:introduction}
Large language models (LLMs) are now widely used to generate code, supporting tasks that range from code completion to whole-function generation.
However, using such code in real-world development requires developers to inspect it and revise it where necessary.
Prior studies report that LLM-generated code may contain functional errors~\cite{liu2024tosem},
security vulnerabilities~\cite{tihanyi2025emse,tony2025tosem}, code smells~\cite{velasco2025icse,siddiq2024msr}, and maintainability issues~\cite{molison2025esem,blyth2025scam}.
Verifying and revising LLM-generated code also imposes additional effort and cognitive load on developers~\cite{tang2024vlhcc}.
Assessing generated code therefore requires looking beyond whether it passes tests and considering maintainability-related quality, such as code smells and structural complexity.

Maintainability-related quality is typically assessed by applying static analysis to the code after it has been generated.
However, code generation itself incurs computational cost and latency, and the generated code may still be functionally incorrect or difficult to maintain.
If such code is discarded after generation and evaluation, both generation and evaluation costs are wasted.
To reduce unnecessary code generation, prior work has used internal representations available before or during generation to predict properties of generated code.
For example, such internal representations have been used to predict functional correctness~\cite{lugoloobi2026lit,varshney2026arxiv,bui2025jss} and compilability or vulnerabilities~\cite{vu2026jss}.
However, these studies mainly target functional correctness or security, and whether internal representations can predict maintainability-related quality before generation remains unexplored.

\begin{figure*}[t]
    \centering
    \includegraphics[width=.8\textwidth]{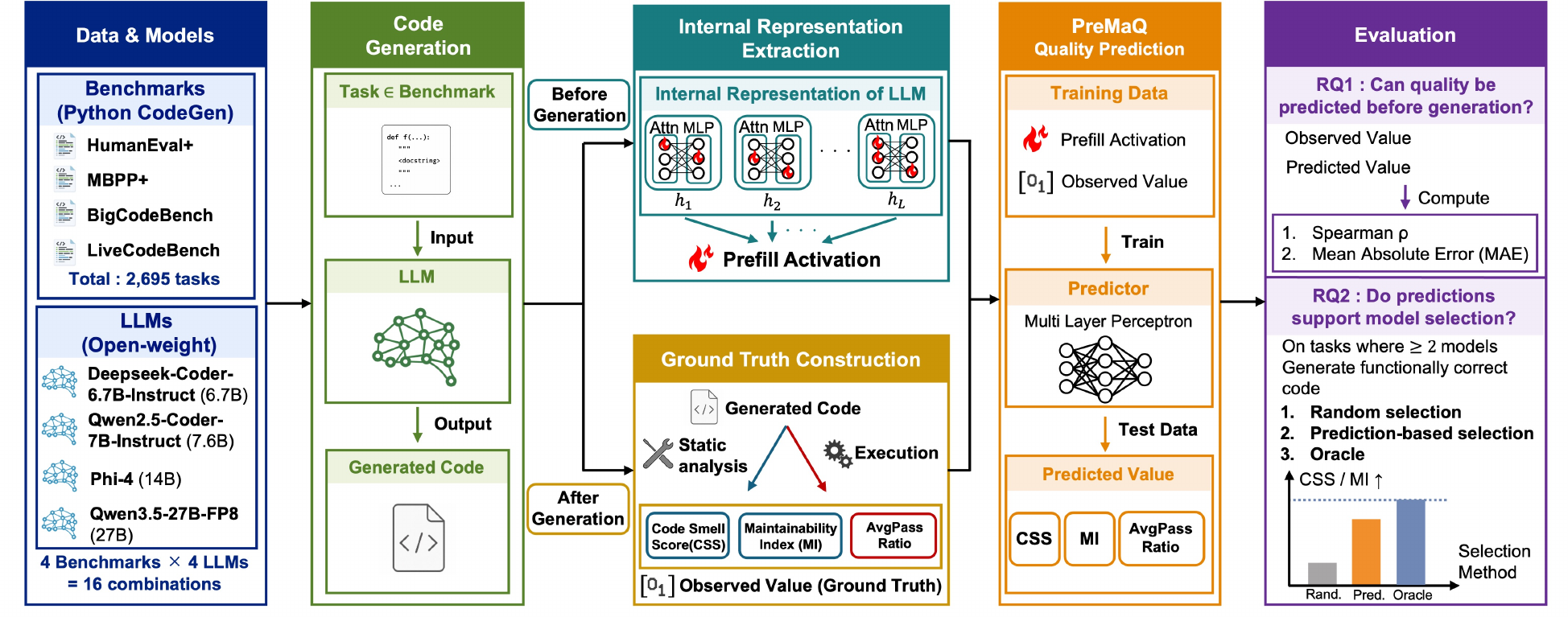}
    \caption{Overview of the PreMaQ workflow and its evaluation.}
    \label{fig:study_overview}
\end{figure*}

Motivated by this open question, this study introduces and evaluates \emph{Pre-Generation Maintainability-Related Quality Prediction (PreMaQ)}, which uses \emph{prefill activations} to predict maintainability-related quality before code generation.
Prefill activations are the internal representations computed while an LLM processes the input prompt.
Figure~\ref{fig:study_overview} shows the PreMaQ workflow.
As the maintainability-related quality to be predicted, we use the following two indicators used in a prior study~\cite{kuang2025arxiv}: Code Smell Score (CSS)~\cite{blyth2025scam,kim2024ase} and Maintainability Index (MI)~\cite{oman1994jss,hericko2023applsci}.

We evaluate prediction both within and across benchmarks to examine whether it extends beyond benchmark-specific patterns.
As a downstream analysis, we examine the \emph{model selection problem} of identifying, among multiple models known to generate functionally correct code for a task, the model likely to produce code with the highest maintainability-related quality.
Our evaluation uses four open-weight LLMs (DeepSeek-Coder-6.7B-Instruct, Qwen2.5-Coder-7B-Instruct, Phi-4, and Qwen3.5-27B-FP8) and $2{,}695$ tasks drawn from four Python code generation benchmarks (HumanEval+, MBPP+, BigCodeBench, and LiveCodeBench).

The main contributions of this study are as follows:
\begin{itemize}[leftmargin=*, itemsep=0pt]
    \item \textbf{PreMaQ predicts maintainability-related quality before code generation.}
    Within benchmarks, its predictions are significantly correlated with observed CSS and MI in all 16 model--benchmark combinations, with mean Spearman $\rho$ values of $0.57$ and $0.65$, respectively. They also generalize across benchmarks for most combinations.
    \item \textbf{PreMaQ supports maintainability-aware model selection.}
    Among models that generate functionally correct code for a task, PreMaQ closes $59.5$\,\% of the quality gap between random and oracle selection, outperforming strategies without internal representations.
    \item \textbf{PreMaQ provides signals complementary to prompt embeddings.}
    Combining predictions from PreMaQ and prompt embeddings closes $60.7$\,\% of the quality gap between random and oracle selection, outperforming either method alone.
\end{itemize}

\section{Motivation}
\label{sec:motivation}
This section reviews prior work to position our work, and then reports a preliminary study that motivates the need for predicting maintainability-related code quality and selecting models likely to produce maintainable code accordingly.

\subsection{Related Work}
\label{sec:related-work}

\noindent\textbf{Analysis of LLM Internal Representations.}
LLM internal representations refer to hidden states or activations computed during prompt processing or token generation.
They include the outputs of the attention and MLP modules and the residual stream, computed at each layer for each token position.
Because these representations reflect how an LLM processes a given input, prior work has used them to infer properties of the LLM's behavior and outputs.
Azaria and Mitchell~\cite{azaria2023emnlp} showed that the truthfulness of LLM output can be inferred from internal representations.
Zhou et al.~\cite{zhou2025icse} detected jailbreak attacks using a neuron activation coverage metric defined over internal representations.

\noindent\textbf{Prediction of Output Quality.}
Because generating output with an LLM incurs computational and financial costs, prior work has sought to estimate output quality before it is produced.
Some methods predict the output quality of each LLM from the input or previous responses and select a model for the task, and others escalate to a more capable model when adequate quality is not expected~\cite{chen2024tmlr,ong2025iclr}.
For code completion, prior work has predicted reference-based completion similarity from the prompt~\cite{sun2025tosem}.
For code generation, internal representations have been used to assess properties of generated code, including functional correctness~\cite{lugoloobi2026lit,varshney2026arxiv,bui2025jss,vu2026jss,ribeiro2026icse,he2026arxiv}, compilability~\cite{vu2026jss}, and the presence of vulnerabilities~\cite{vu2026jss,sternfeld2026arxiv}.
However, the predicted properties are mainly functional correctness or security, and whether maintainability-related quality indicators, such as CSS and MI, can be predicted before generation remains unexplored.
This study therefore investigates whether these indicators can be predicted before generation from internal representations.

\subsection{Preliminary Study}
\label{sec:preliminary}

To motivate predicting maintainability-related quality directly, we set the following Preliminary Question (PQ).
\begin{enumerate}[label=\textit{PQ}, leftmargin=*, align=left, itemsep=0pt]
    \item \textbf{To what extent model size or functional correctness can identify the model that generates code with the highest maintainability-related quality.}
\end{enumerate}

If readily available information, such as model size, or existing predictions of functional correctness were sufficient to identify such a model, there would be little need to predict maintainability-related quality directly.
To answer the PQ, we conduct a preliminary study using the same experimental setting as the main study: four local LLMs, four code generation benchmarks, two maintainability-related quality indicators (CSS and MI), and one functional correctness indicator (AvgPassRatio).
Higher values are better for all three indicators.

\begin{table*}[t]
\centering
\caption{Functional correctness and maintainability-related quality of the four models.}
\label{tab:prelim_local}
\resizebox{0.9\linewidth}{!}{%
\begin{tabular}{@{}lll ccccc@{}}
\toprule
Metric & Model & \#Params. & HumanEval+ & MBPP+ & BigCodeBench & LiveCodeBench & Average \\
\midrule
  \multirow{4}{*}{AvgPassRatio} & DeepSeek-Coder-6.7B & 6.7B & 0.833 & 0.630 & 0.692 & 0.267 & 0.525 \\
  & Qwen2.5-Coder-7B & 7B & 0.919 & 0.828 & 0.737 & 0.420 & 0.637 \\
  & Phi-4 & 14B & 0.910 & 0.740 & 0.775 & 0.427 & 0.642 \\
  & Qwen3.5-27B-FP8 & 27B & \textbf{0.973} & \textbf{0.839} & \textbf{0.814} & \textbf{0.796} & \textbf{0.820} \\
\midrule
  \multirow{4}{*}{\shortstack[l]{CSS\\all / pass}} & DeepSeek-Coder-6.7B & 6.7B & 4.098 / 4.011 & 2.581 / 1.914 & \textbf{5.597} / \textbf{5.417} & 3.980 / 3.621 & \textbf{4.450} / \textbf{4.207} \\
  & Qwen2.5-Coder-7B & 7B & 4.370 / 4.212 & 1.248 / 1.038 & 4.005 / 3.937 & \textbf{4.371} / 3.597 & 3.784 / 3.274 \\
  & Phi-4 & 14B & 4.401 / 4.245 & 3.181 / 2.744 & 4.440 / 4.337 & 3.459 / \textbf{3.639} & 3.877 / 3.867 \\
  & Qwen3.5-27B-FP8 & 27B & \textbf{6.096} / \textbf{5.991} & \textbf{4.365} / \textbf{4.071} & 4.927 / 4.854 & 2.131 / 2.532 & 3.825 / 3.869 \\
\midrule
  \multirow{4}{*}{\shortstack[l]{MI\\all / pass}} & DeepSeek-Coder-6.7B & 6.7B & \textbf{74.04} / \textbf{74.87} & 79.07 / \textbf{83.69} & \textbf{77.62} / \textbf{80.12} & \textbf{60.91} / \textbf{69.72} & \textbf{71.06} / \textbf{78.24} \\
  & Qwen2.5-Coder-7B & 7B & 72.46 / 73.20 & \textbf{80.87} / 82.63 & 77.15 / 78.71 & 60.22 / 66.88 & 70.76 / 76.10 \\
  & Phi-4 & 14B & 71.39 / 72.35 & 76.57 / 79.98 & 75.52 / 76.50 & 54.99 / 65.39 & 67.38 / 74.05 \\
  & Qwen3.5-27B-FP8 & 27B & 68.97 / 69.29 & 76.86 / 78.37 & 71.16 / 72.86 & 54.37 / 55.60 & 65.25 / 66.24 \\
\bottomrule
\end{tabular}%
}
\end{table*}

\textbf{Model size is not a reliable predictor of maintainability-related quality.}
\tabref{tab:prelim_local} shows the quality of the four models. Average is the task-count-weighted mean over the four benchmarks.
CSS and MI are shown as ``all / pass,'' computed over all generated outputs and over functionally correct outputs only, respectively.
Bold indicates the best-performing model in each column.
Functional correctness increases with model size.
The average AvgPassRatio monotonically increases from $0.525$ for the 6.7B model to $0.820$ for the 27B model.
However, the average maintainability-related quality does not necessarily improve with model size.
These results indicate that selecting the largest model is not sufficient when the goal is to obtain code with high maintainability-related quality.

\begin{table}[t]
\centering
\caption{Distribution of $k$, the number of LLMs that solved each task.}
\label{tab:kdist}
\resizebox{\columnwidth}{!}{%
\begin{tabular}{@{}l r ccccc r r@{}}
\toprule
Dataset & Total & $k{=}0$ & $k{=}1$ & $k{=}2$ & $k{=}3$ & $k{=}4$ & $k{\geq}1$ & $k{\geq}2$ \\
\midrule
  HumanEval+ & 164 & 10 & 10 & 13 & 28 & 103 & 93.9\% & 87.8\% \\
  MBPP+ & 378 & 58 & 31 & 63 & 73 & 153 & 84.7\% & 76.5\% \\
  BigCodeBench & 1098 & 275 & 140 & 147 & 213 & 323 & 75.0\% & 62.2\% \\
  LiveCodeBench & 1055 & 259 & 378 & 161 & 130 & 127 & 75.5\% & 39.6\% \\
\midrule
  Overall & 2695 & 602 & 559 & 384 & 444 & 706 & 77.7\% & 56.9\% \\
\bottomrule
\end{tabular}%
}
\end{table}

\textbf{Functional correctness alone cannot identify the model with the highest maintainability-related quality.}
We examine whether functional correctness is sufficient for selecting models that generate highly maintainable code.
\tabref{tab:kdist} shows how many of the four LLMs solved each task, denoted by $k$; the last two columns show the percentages of tasks with $k{\geq}1$ and $k{\geq}2$.
Among the $2{,}695$ code generation tasks in the four benchmarks, $56.9$\,\% are solved by two or more models.
This means that, for the majority of tasks, predicting functional correctness leaves multiple candidate models.
However, these candidates differ substantially in maintainability-related quality: for each task solved by two or more models, we compute the gap in each indicator between the best and worst functionally correct candidates.
The average gap is $2.83$ for CSS on a 10-point scale and $8.49$ for MI on a 100-point scale.

\summaryblock{Motivation for predicting maintainability-related quality}{%
Neither model size nor functional correctness identifies the model that generates the code with the highest maintainability-related quality.
Substantial differences remain even among functionally correct candidates.
Predicting this quality before generation therefore has value for selecting higher-quality code.
}

\section{PreMaQ}
\label{sec:method}

This section describes PreMaQ, our approach for predicting the quality of generated code from prefill activations before generation begins.
We first formulate the problem and then describe the internal representations used as features, the quality indicators to be predicted, and the predictor architecture.

\subsection{Problem Formulation}
\label{subsec:method_problem}

PreMaQ predicts the maintainability-related quality of the code that an LLM will generate, using the internal representations computed while the LLM processes the input prompt.
Hereafter, we refer to these representations as \textit{prefill activations}.
Let $\mathcal{D}$ be a set of input prompts in a code generation benchmark, and let $f$ be an LLM that generates code.
For a prompt $x \in \mathcal{D}$, we denote the code generated by $f$ as $f(x)$.
For a quality indicator $q$, the quality of this generated code is denoted by $q(f(x))$.
The code quality predictor $g^{\mathcal{D}}_{f,q}$ takes as input the feature vector $z_f(x)$ derived from the prefill activations obtained by forwarding $x$ through $f$, and predicts the value of the quality indicator $q(f(x))$.
Because $z_f(x)$ can be obtained with only one forward pass, the quality can be estimated without paying the cost of code generation.
Since the dimensionality and distribution of $z_f(x)$ differ across LLMs, we train an independent predictor $g^{\mathcal{D}}_{f,q}$ for each combination of benchmark $\mathcal{D}$, LLM $f$, and quality indicator $q$.

\subsection{Prefill Activations and Features}
\label{subsec:method_features}

As prefill activations, we use neuron outputs at the final token position immediately after the LLM $f$ processes the prompt $x$.
In autoregressive LLMs, because of the causal attention structure, only the final token position can attend to all tokens in the prompt.
Thus, the neuron outputs at the final token position reflect the internal representation after the model $f$ has processed the entire prompt $x$; prior work has also used internal representations from this position~\cite{azaria2023emnlp,zhou2025icse}.
We therefore use this final-token position to obtain representations after the whole prompt has been processed.

Let $L_f$ be the number of stacked Transformer blocks in $f$, and let $d_f$ be the dimensionality of its hidden states.
Let $h_\ell(x) \in \mathbb{R}^{d_f}$ be the vector formed by the neuron outputs extracted from layer $\ell$ ($1 \leq \ell \leq L_f$) immediately after the final token of prompt $x$ is processed.
We construct a feature vector by concatenating the vectors from all layers:
\begin{equation*}
a_f(x)
=
\operatorname{concat}\!\left(
h_1(x), h_2(x), \ldots, h_{L_f}(x)
\right)
\in \mathbb{R}^{d_fL_f}.
\end{equation*}

For the source of $h_\ell(x)$, we compare four activation sources: the output of the Attention layer in the $\ell$-th block, the output of the MLP layer, the concatenation of the Attention and MLP outputs, and the residual stream after passing through the block.

We standardize each dimension of $a_f(x)$ using z-score normalization and denote the resulting feature vector by
$z_f(x) \in \mathbb{R}^{d_f L_f}$.
This normalization is needed because the scale of activations can vary substantially across layers and dimensions.
The $k$-th dimension of $z_f(x)$ is defined as
\begin{equation*}
z_{f,k}(x)
=
\frac{a_{f,k}(x)-\mu_{f,k}}{\sigma_{f,k}},
\qquad
1 \leq k \leq d_f L_f.
\end{equation*}
Here, $a_{f,k}(x)$ denotes the $k$-th dimension of $a_f(x)$ for prompt $x$, and $\mu_{f,k}$ and $\sigma_{f,k}$ are the mean and standard deviation of $a_{f,k}(x)$ across all prompts in the training data.
\subsection{Quality Indicators to Be Predicted}
\label{subsec:metrics}

We describe the quality indicators used as prediction targets in this study.
Following Kuang et al.~\cite{kuang2025arxiv}, we choose indicators from two perspectives: the maintainability-related quality that this study focuses on, and functional correctness, which has been widely predicted in prior work~\cite{bui2025jss,vu2026jss,lugoloobi2026lit} and serves as a comparison target.

\noindent\textbf{Functional Correctness.}
We evaluate and predict functional correctness using AvgPassRatio, as adopted in prior code-generation studies~\cite{kuang2025arxiv,hao2022arxiv,dong2025tosem}.
It is the fraction of test cases passed by the generated code.
Let $N_t$ be the total number of test cases associated with a task, and let $N_p$ be the number of test cases passed by the generated code.
AvgPassRatio is calculated as $N_p/N_t$.
Its range is $[0,1]$, and a higher value indicates better code generation performance in terms of functional correctness.

\noindent\textbf{Maintainability-Related Quality.}
We evaluate and predict maintainability-related quality using Code Smell Score (CSS) and Maintainability Index (MI).

CSS~\cite{kuang2025arxiv,blyth2025scam,kim2024ase} is a code quality indicator based on static analysis.
It is computed by weighting the violations detected by Pylint~\cite{pylint2026webpage} by message type, normalizing their weighted sum by the number of statements in the code, and subtracting the result from the maximum score:
\begin{equation*}
\resizebox{\columnwidth}{!}{$\displaystyle
\mathrm{CSS} = \max\!\left(0,\; 10 - 10 \times \frac{w_e N_e + w_w N_w + w_r N_r + w_c N_c}{N_s}\right).
$}
\end{equation*}
Its range is $[0,10]$, and a higher value indicates fewer code smells.
Here, $N_e$, $N_w$, $N_r$, and $N_c$ are the numbers of Error, Warning, Refactor, and Convention messages, respectively, and $N_s$ is the number of statements in the code.
Following prior work~\cite{kuang2025arxiv}, we set $w_e = 5.0$ and $w_w = w_r = w_c = 1.0$.
If Pylint emits a Fatal message, we set $\mathrm{CSS}=0$.

MI~\cite{kuang2025arxiv,oman1994jss,hericko2023applsci} is a maintainability indicator that summarizes how easy software is to maintain in a single value.
It is computed from Halstead Volume ($HV$), which captures lexical complexity and program size, cyclomatic complexity ($CC$), which captures control-flow complexity, and lines of code ($LOC$):
\begin{equation*}
\resizebox{\columnwidth}{!}{$\displaystyle
\mathrm{MI} = \max\!\left(0,\; 100 \times \frac{171 - w_{HV}\ln(HV) - w_{CC}\,CC - w_{LOC}\ln(LOC)}{171}\right)
$}
\end{equation*}
Its range is $[0,100]$, and a higher value indicates better maintainability.
The weights $w_{HV}$, $w_{CC}$, and $w_{LOC}$ represent the contribution of each component.
Following prior work~\cite{kuang2025arxiv}, we set $w_{HV} = 5.2$, $w_{CC} = 0.23$, and $w_{LOC} = 16.2$.
We compute $HV$, $CC$, and $LOC$ using radon~\cite{lacchia2026webpage}.

\subsection{Predictor Architecture}
\label{subsec:method_arch}

Following prior work that predicts properties of LLM outputs from internal representations using multilayer perceptrons (MLPs)~\cite{azaria2023emnlp,zhou2025icse,bui2025jss,vu2026jss,varshney2026arxiv}, we instantiate the predictor $g^{\mathcal{D}}_{f,q}$ as an MLP.
In this study, the predictor serves as a probe for measuring whether the internal representation $z_f(x)$ contains information about the quality of the code to be generated.
The MLP $g^{\mathcal{D}}_{f,q}$ maps $z_f(x)$ to a predicted value of $q(f(x))$.
Because our primary purpose is to assess the quality-related information contained in $z_f(x)$ rather than to compare predictor architectures, we use the MLP architecture adopted by Zhou et al.~\cite{zhou2025icse}.
We train $g^{\mathcal{D}}_{f,q}$ as a regression model because all prediction targets, AvgPassRatio, CSS, and MI, are continuous values.

\section{Study Design}
\label{sec:study-design}

This section presents the research questions, subject models and benchmarks, data splits, predictor training and evaluation settings for RQ1, model selection procedure and evaluation for RQ2, and implementation details.

\subsection{Research Questions}
\label{subsec:problem}

We formulate the following research questions to examine prediction performance and its application to model selection.

\begin{enumerate}[label=\textit{RQ\arabic*}, leftmargin=*, align=left, itemsep=0pt]
    \item \textbf{How well does PreMaQ predict the maintainability-related quality of generated code?}
    \item \textbf{How does PreMaQ-based model selection compare with selection without internal representations in terms of maintainability-related quality?}
\end{enumerate}

RQ1 evaluates PreMaQ's prediction performance within and across benchmarks for each LLM--benchmark combination, using functional correctness as a comparison target and comparing four activation sources.
RQ2 evaluates whether PreMaQ-based model selection improves maintainability-related quality over strategies without internal representations.
To isolate maintainability-related quality, we restrict the candidates for each task to models known to generate functionally correct code.
This setting with known functionally correct candidates builds on prior work on pre-generation prediction of functional correctness~\cite{lugoloobi2026lit,varshney2026arxiv}.

\begin{table}[t]
    \centering
    \caption{Overview of the experimented models.}
    \label{tab:subject_llm}
    \resizebox{1.0\linewidth}{!}{
    \begin{tabular}{lrll}
        \toprule
        Model & \#Params. & Type & Attention \\
        \midrule
        DeepSeek-Coder-6.7B-Instruct~\cite{guo2024arxiv} & 6.7B & Code-specialized & Self-attention \\
        Qwen2.5-Coder-7B-Instruct~\cite{hui2024arxiv} & 7.6B & Code-specialized & Self-attention \\
        Phi-4~\cite{abdin2024arxiv} & 14.7B & General-purpose & Self-attention \\
        Qwen3.5-27B-FP8~\cite{qwen2026webpage} & 27.0B & General-purpose & Hybrid attention \\
        \bottomrule
    \end{tabular}
    }
\end{table}

\subsection{Experimented Large Language Models}
\label{subsec:models}

\tabref{tab:subject_llm} summarizes the models experimented in this study.
We use four open-weight LLMs: DeepSeek-Coder-6.7B-Instruct~\cite{guo2024arxiv}, Qwen2.5-Coder-7B-Instruct~\cite{hui2024arxiv}, Phi-4~\cite{abdin2024arxiv}, and Qwen3.5-27B-FP8~\cite{qwen2026webpage}.
We focus on open-weight models because their internal representations can be obtained.
To reduce dependence on specific model characteristics, we include both code-specialized and general-purpose models, with sizes ranging from 6.7B to 27B parameters.
We also include a model that uses hybrid attention, in addition to models based on self-attention (\tabref{tab:subject_llm}).
Recent LLMs have adopted increasingly diverse architectures.
Including this model broadens the architectural variation among the subject models and allows us to examine whether prefill activations are useful beyond models with a single attention architecture.
We use decoder-only models, an architecture commonly used for code generation~\cite{zan2023acl}.

\subsection{Benchmarks}
\label{subsec:benchmarks}

We use four representative Python code generation benchmarks: HumanEval+~\cite{chen2021arxiv,liu2023nips}, MBPP+~\cite{austin2021arxiv,liu2023nips}, BigCodeBench~\cite{zhuo2025iclr}, and LiveCodeBench~\cite{jain2025iclr}.
Each benchmark consists of tasks that take natural-language problem descriptions as input and require the model to generate the corresponding programs.
Each task also includes multiple test cases for judging whether the generated code works correctly.

For BigCodeBench, we exclude 42 tasks whose benchmark-provided canonical solutions do not pass the accompanying tests in our execution environment, and use the remaining 1,098 tasks.
For LiveCodeBench, we use dataset version v6.
As a result, HumanEval+, MBPP+, BigCodeBench, and LiveCodeBench contain 164, 378, 1,098, and 1,055 code generation tasks, respectively, for a total of 2,695 tasks.

\subsection{Data Splits for Training and Evaluation}
\label{subsec:method_split}

For each LLM--quality-indicator combination, we train and evaluate $g^{\mathcal{D}}_{f,q}$ in two settings: a within-benchmark setting, which uses data from the same benchmark, and a cross-benchmark setting, which uses data from different benchmarks.
In each setting, let $\mathcal{D}_{\mathrm{train}}$ be the set of prompts used to train $g^{\mathcal{D}_{\mathrm{train}}}_{f,q}$ and let $\mathcal{D}_{\mathrm{eval}}$ be the set of prompts used for evaluation.
For the within-benchmark setting, we use 5-fold cross validation.
For the cross-benchmark setting, we train the predictor on three of the four benchmarks and evaluate it on the remaining benchmark.
Standardization statistics are computed only from the training folds.

\subsection{Model Training Setting}
\label{subsec:training}

The prediction targets, AvgPassRatio, CSS, and MI, are all learned as regression targets, and the loss function is mean squared error (MSE).
We use MSE as the training objective because it is a standard smooth loss for continuous regression.
During training, we normalize all prediction targets to the range $[0,1]$.
For the 5-fold cross validation, we fix the random seed for fold assignment and obtain out-of-fold predictions so that every task is included in the evaluation data exactly once.
These settings are shared across all combinations of LLM, benchmark, prediction target, and input feature.
Only the input features and labels differ across combinations.

\subsection{Evaluation Metrics for Prediction Performance}
\label{subsec:method_metrics}

We evaluate prediction performance by comparing the predicted value with the observed value of each quality indicator.
The observed value $q(f(x))$ is obtained by actually generating code with the LLM.
We consider two perspectives: (1) absolute error between predicted and observed values, and (2) correlation between predicted and observed values.

For absolute error, we use Mean Absolute Error (MAE):
\begin{equation*}
\mathrm{MAE}(\mathcal{D}_{\mathrm{eval}}) =
\frac{1}{|\mathcal{D}_{\mathrm{eval}}|}
\sum_{x \in \mathcal{D}_{\mathrm{eval}}}
\bigl| g^{\mathcal{D}_{\mathrm{train}}}_{f,q}\bigl(z_f(x)\bigr) - q\bigl(f(x)\bigr) \bigr|.
\end{equation*}
We use MAE rather than MSE because it is less dominated by a small number of large individual errors.

For correlation between predicted and observed values, we use Spearman's rank correlation coefficient $\rho$ because the observed target values did not satisfy normality according to the Shapiro--Wilk test.

\subsection{Model Selection Procedure for RQ2}
\label{subsec:selection}

In RQ2, we apply the predictors constructed in RQ1 to model selection for code generation on each task.

\begin{figure*}[t]
    \centering
    \includegraphics[width=.8\textwidth]{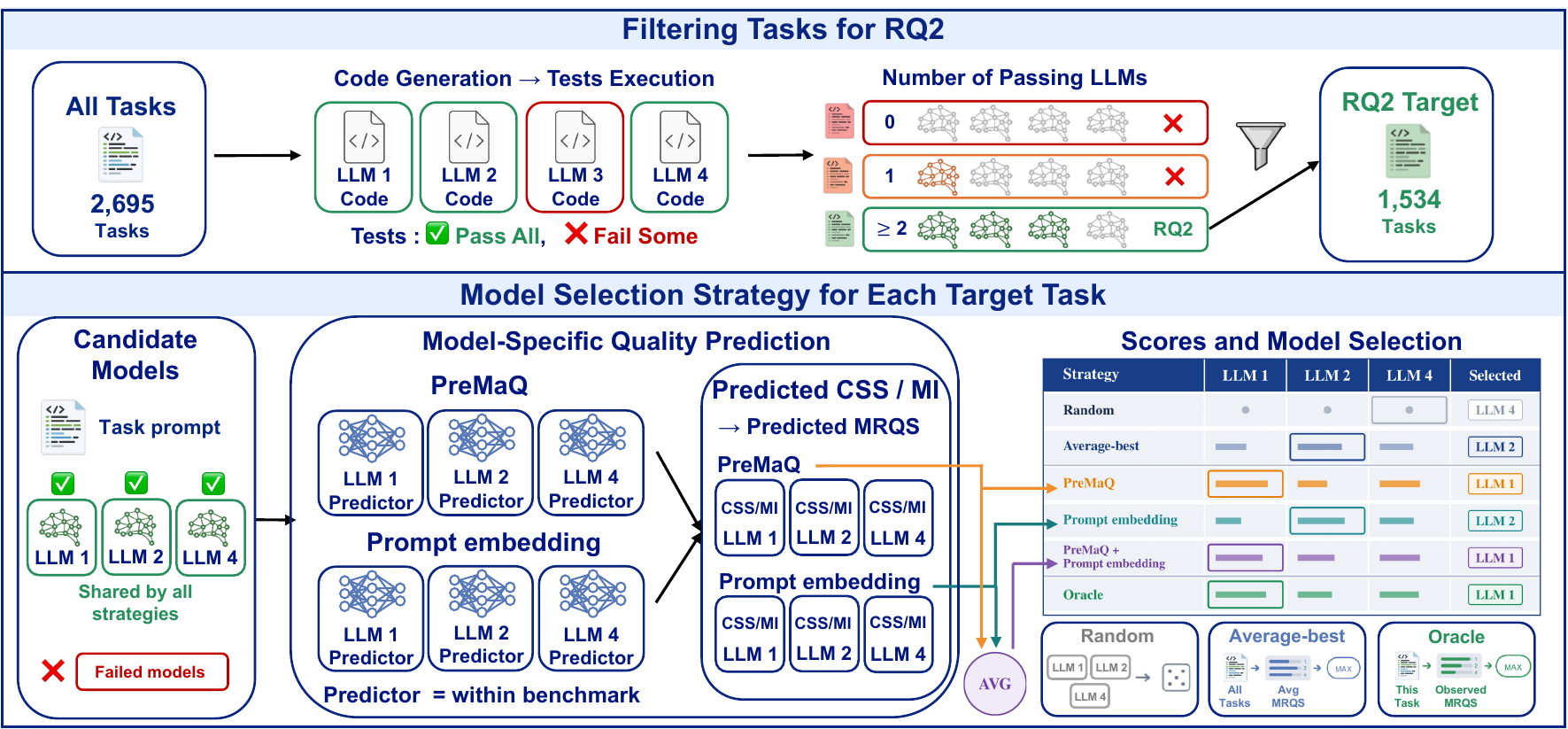}
    \caption{Overview of the task filtering process and model selection strategies used in RQ2.}
    \label{fig:rq2_strategy}
\end{figure*}

\noindent\textbf{Model Selection Problem.}
RQ2 addresses the problem of selecting the model likely to generate code with the highest maintainability-related quality among functionally correct candidates.
We restrict the analysis to tasks for which at least two of the four models generated functionally correct code.
For each such task, we assume that the functionally correct models are known and use them as candidates for selection.
The upper part of \figref{fig:rq2_strategy} illustrates this task filtering process.
This setting builds on prior work on pre-generation prediction of whether a model can solve a task~\cite{lugoloobi2026lit,varshney2026arxiv}.

\noindent\textbf{Maintainability-Related Quality Indicator for Selection.}
To rank candidate models by a single criterion, we use a score that rewards both fewer code smells and higher maintainability.
Specifically, we normalize CSS and MI to $[0,1]$ as $c=\mathrm{CSS}/10$ and $m=\mathrm{MI}/100$, and define the Maintainability-Related Quality Score (MRQS) as their harmonic mean:
$
\mathrm{MRQS} = 2cm/(c + m),
$
where $\mathrm{MRQS}=0$ when $c+m=0$.

\noindent\textbf{Selection Procedure.}
For each task, we select the candidate model with the highest predicted MRQS and record the observed MRQS of its generated code.
For the prediction-based strategies, each task is scored using out-of-fold predictions, so the predictor used for that task is not trained on the task itself.

\noindent\textbf{Selection Strategies.}

We compare six selection strategies on the same target tasks, covering a lower bound, baselines, our proposed white-box approach, a combination strategy, and an upper bound.
The lower part of \figref{fig:rq2_strategy} illustrates the strategies.
\begin{enumerate}[label=(\arabic*), leftmargin=*, align=left, itemsep=0pt]
    \item \textbf{Random (lower bound)}: randomly selects one candidate with equal probability, serving as a lower bound.
    \item \textbf{Average-best (baseline)}: selects the candidate model with the highest mean observed MRQS across all tasks in the four benchmarks. This is a fixed, post-hoc reference that does not use task-specific information.
    \item \textbf{Prompt embedding (black-box baseline)}~\cite{reimers2019emnlp}: embeds the prompt using an external embedding model, all-MiniLM-L6-v2 from sentence-transformers, and trains predictors on these embeddings using the same architecture and data splits as the predictors based on PreMaQ. It selects the candidate model with the highest predicted MRQS. This is a black-box baseline because it does not access the internal representations of the candidate models. This strategy follows prior work in predicting output quality from prompt-derived information~\cite{sun2025tosem,ong2025iclr}. 
    \item \textbf{PreMaQ (proposed, white-box)}: our proposed approach, which selects the candidate model with the highest MRQS predicted from prefill activations. This is a white-box approach because it requires access to the internal representations of the candidate models.
    \item \textbf{PreMaQ + prompt embedding (combination)}: selects the candidate model with the highest average of the MRQS values predicted from PreMaQ and prompt embeddings. This strategy is included to examine whether the two signals provide complementary information for model selection.
    \item \textbf{Oracle (upper bound)}: selects the model with the highest observed MRQS, serving as the achievable upper bound.
\end{enumerate}

\noindent\textbf{Evaluation of Selection Strategies.}
We evaluate each strategy by averaging the observed MRQS of the code generated by its selected models across the target tasks.
For Random, we report the expected mean MRQS over all possible selections, avoiding dependence on a random seed.
We also report the relative gain, which measures the fraction of the gap between Random and Oracle closed by each strategy.
Let $\bar{q}_s$ denote the mean MRQS achieved by strategy $s$; its relative gain is
\begin{equation*}
    \mathrm{RelGain}_q(s) =
    \frac{\bar{q}_s-\bar{q}_{\mathrm{Random}}}
         {\bar{q}_{\mathrm{Oracle}}-\bar{q}_{\mathrm{Random}}}.
\end{equation*}

\subsection{Implementation Details}
\label{subsec:env}

All experiments were run on a machine equipped with an Intel(R) Xeon(R) w7-2595X CPU (26 cores, up to 4.8 GHz) and an NVIDIA RTX 6000 Ada Generation GPU.
Detailed version information for all libraries is available in our replication package.

\section{Results}
\label{sec:results}

In our environment, training one predictor took approximately 4 to 33 seconds, depending on the feature dimensionality and the number of tasks.
Quality prediction with a trained predictor took less than 0.4 ms per task.
The additional computation required for quality prediction is only one forward pass for obtaining prefill activations and the predictor inference, making it much faster than generating code and measuring its quality afterward.

\subsection{RQ1: Quality Prediction based on PreMaQ}
\label{subsec:rq1}

\subsubsection{Effective Activation Source}
\tabref{tab:rq1_layertype} reports, for each combination of the four activation sources and three targets, the task-count-weighted mean Spearman correlation between predicted and observed values across model--benchmark combinations.
The best value in each row is bolded.

\begin{table}[t]
\centering
\caption{Within-benchmark mean Spearman correlations across four activation sources.}
\label{tab:rq1_layertype}
\begin{tabular}{@{}lcccc@{}}
\toprule
Target & Attention & MLP & Attn+MLP & Residual \\
\midrule
  \begin{tabular}[c]{@{}l@{}}AvgPassRatio ($\rho$)\end{tabular} & \textbf{0.3236} & 0.3202 & 0.3167 & 0.3160 \\
  CSS ($\rho$) & \textbf{0.5728} & 0.5574 & 0.5537 & 0.5591 \\
  MI ($\rho$) & \textbf{0.6506} & 0.6466 & 0.6404 & 0.6474 \\
\bottomrule
\end{tabular}
\end{table}

\textbf{Attention outputs yield the highest correlations with observed values.}
Attention outputs achieve the highest $\rho$ for all three prediction targets.
Therefore, we regard Attention outputs as the most effective source for pre-generation code quality prediction.
Accordingly, all subsequent analyses use Attention outputs as the activation source.

\subsubsection{Within-Benchmark Setting}
\tabref{tab:rq1_quality} summarizes the within-benchmark results.
Each value before parentheses is a task-count-weighted mean over the model--benchmark combinations for each row.
The percentage in parentheses denotes the proportion of model--benchmark combinations in each row that are statistically significant at the 5\% level after Holm correction.
For $\rho$, significance indicates rejection of the null hypothesis of no correlation.
For MAE, we compare PreMaQ with a mean-value baseline, a constant predictor that always outputs the mean observed value in the training data.
Significance indicates that PreMaQ achieves lower MAE than this baseline according to a one-sided paired permutation test.

\begin{table}[t]
\centering
\caption{Within-benchmark prediction performance.}
\label{tab:rq1_quality}
\resizebox{.9\linewidth}{!}{%
\begin{tabular}{@{}lcc@{}}
\toprule
Quality indicator & $\rho$ (\%sig.) & MAE (\%sig.) \\
\midrule
AvgPassRatio & 0.32 (75\%) & 0.25 (88\%) \\
CSS & 0.57 (100\%) & 1.56 (100\%) \\
MI & 0.65 (100\%) & 8.1 (100\%) \\
\midrule
Model & CSS $\rho$ (\%sig.) & MI $\rho$ (\%sig.) \\
\midrule
DeepSeek-Coder-6.7B & 0.56 (100\%) & 0.61 (100\%) \\
Qwen2.5-Coder-7B & 0.64 (100\%) & 0.71 (100\%) \\
Phi-4 & 0.48 (100\%) & 0.65 (100\%) \\
Qwen3.5-27B-FP8 & 0.61 (100\%) & 0.62 (100\%) \\
\bottomrule
\end{tabular}%
}
\end{table}

\textbf{Within benchmarks, PreMaQ yields significant correlations and lower MAE than the mean-value baseline.}
As shown in \tabref{tab:rq1_quality}, predicted CSS and MI values are significantly correlated with their observed values for every model--benchmark combination (100\%).
The mean correlations across the 16 model--benchmark combinations are $\rho{=}0.57$ for CSS and $\rho{=}0.65$ for MI, approximately twice the mean correlation for AvgPassRatio.
The median values also show the same ordering of prediction targets: $0.24$ for AvgPassRatio, $0.51$ for CSS, and $0.64$ for MI.
The MAE of PreMaQ is significantly lower than that of the mean-value baseline in all combinations for both of CSS and MI.
These results indicate that, by using prefill activations, PreMaQ predicts maintainability-related quality more consistently than it predicts functional correctness.

\subsubsection{Cross-Benchmark Setting}
\tabref{tab:rq1_crossdataset} summarizes the cross-benchmark results alongside the corresponding within-benchmark results.
$^{*}$ denotes statistical significance at the 5\% level after Holm correction.
Statistical significance is assessed using the same tests as in the within-benchmark setting.
The final two rows report the task-count-weighted mean and the median across the 16 model--benchmark combinations.

\begin{table}[t]
\centering
\caption{Cross-benchmark generalization performance. Each cell reports Within/Cross results. HE+, BCB, and LCB denote HumanEval+, BigCodeBench, and LiveCodeBench.}
\label{tab:rq1_crossdataset}
\resizebox{\linewidth}{!}{%
\begin{tabular}{@{}ll cc cc@{}}
\toprule
 & & \multicolumn{2}{c}{CSS} & \multicolumn{2}{c}{MI} \\
\cmidrule(lr){3-4}\cmidrule(lr){5-6}
Model & Bench. & $\rho$ & MAE & $\rho$ & MAE \\
\midrule
  \multirow{4}{*}{\makecell[l]{DeepSeek-\\Coder-6.7B}} & HE+ & 0.40$^{*}$/0.38$^{*}$ & 2.14$^{*}$/2.40 & 0.45$^{*}$/0.62$^{*}$ & 8.2$^{*}$/8.0 \\
   & MBPP+ & 0.49$^{*}$/0.45$^{*}$ & 1.95$^{*}$/2.19$^{*}$ & 0.62$^{*}$/0.62$^{*}$ & 8.1$^{*}$/9.8$^{*}$ \\
   & BCB & 0.51$^{*}$/0.09$^{*}$ & 1.28$^{*}$/1.86$^{*}$ & 0.58$^{*}$/0.37$^{*}$ & 10.4$^{*}$/13.5$^{*}$ \\
   & LCB & 0.66$^{*}$/0.19$^{*}$ & 1.65$^{*}$/5.71 & 0.68$^{*}$/-0.07$^{*}$ & 5.7$^{*}$/41.7 \\
  \cmidrule(l){2-6}
  \multirow{4}{*}{\makecell[l]{Qwen2.5-\\Coder-7B}} & HE+ & 0.37$^{*}$/0.44$^{*}$ & 2.25$^{*}$/2.39 & 0.56$^{*}$/0.66$^{*}$ & 7.5$^{*}$/6.8$^{*}$ \\
   & MBPP+ & 0.50$^{*}$/0.36$^{*}$ & 1.12$^{*}$/1.22$^{*}$ & 0.75$^{*}$/0.74$^{*}$ & 6.3$^{*}$/8.0$^{*}$ \\
   & BCB & 0.57$^{*}$/0.08$^{*}$ & 1.70$^{*}$/2.39 & 0.65$^{*}$/0.51$^{*}$ & 9.6$^{*}$/13.1$^{*}$ \\
   & LCB & 0.81$^{*}$/0.47$^{*}$ & 1.09$^{*}$/2.83$^{*}$ & 0.79$^{*}$/0.69$^{*}$ & 4.5$^{*}$/11.6$^{*}$ \\
  \cmidrule(l){2-6}
  \multirow{4}{*}{Phi-4} & HE+ & 0.42$^{*}$/0.38$^{*}$ & 2.16$^{*}$/2.32 & 0.50$^{*}$/0.70$^{*}$ & 8.7$^{*}$/7.4$^{*}$ \\
   & MBPP+ & 0.43$^{*}$/0.42$^{*}$ & 2.11$^{*}$/2.29 & 0.71$^{*}$/0.66$^{*}$ & 7.7$^{*}$/9.2$^{*}$ \\
   & BCB & 0.49$^{*}$/0.18$^{*}$ & 1.77$^{*}$/2.08$^{*}$ & 0.54$^{*}$/0.47$^{*}$ & 10.8$^{*}$/12.8$^{*}$ \\
   & LCB & 0.50$^{*}$/0.18$^{*}$ & 1.49$^{*}$/3.13 & 0.77$^{*}$/0.67$^{*}$ & 7.2$^{*}$/20.2$^{*}$ \\
  \cmidrule(l){2-6}
  \multirow{4}{*}{\makecell[l]{Qwen3.5-\\27B-FP8}} & HE+ & 0.56$^{*}$/0.63$^{*}$ & 1.54$^{*}$/1.61$^{*}$ & 0.57$^{*}$/0.73$^{*}$ & 7.3$^{*}$/6.6$^{*}$ \\
   & MBPP+ & 0.69$^{*}$/0.66$^{*}$ & 1.69$^{*}$/2.16$^{*}$ & 0.70$^{*}$/0.76$^{*}$ & 8.1$^{*}$/7.2$^{*}$ \\
   & BCB & 0.55$^{*}$/0.21$^{*}$ & 1.66$^{*}$/2.04$^{*}$ & 0.55$^{*}$/0.49$^{*}$ & 10.8$^{*}$/11.1$^{*}$ \\
   & LCB & 0.65$^{*}$/-0.18$^{*}$ & 1.28$^{*}$/4.30 & 0.67$^{*}$/0.63$^{*}$ & 6.8$^{*}$/7.0$^{*}$ \\
  \midrule
  \multicolumn{2}{l}{\textit{Weighted mean}} & 0.57/0.22 & 1.56/2.82 & 0.65/0.52 & 8.1/14.7 \\
  \multicolumn{2}{l}{\textit{Median}} & 0.50/0.37 & 1.68/2.31 & 0.64/0.65 & 7.9/9.5 \\
\bottomrule
\end{tabular}%
}
\end{table}

\textbf{PreMaQ shows cross-benchmark generalizability.}
\tabref{tab:rq1_crossdataset} shows that PreMaQ achieves positive and statistically significant correlations across benchmarks in 15 of 16 model--benchmark combinations for both MI and CSS.
For MI in particular, PreMaQ achieves performance comparable to its within-benchmark performance, exceeding it in some cases when evaluated across benchmarks.
This limited degradation indicates that the predictors do not merely fit benchmark-specific patterns.

\summaryblock{Answer to RQ1}{
Within benchmarks, PreMaQ achieves significant correlations and significantly lower MAE than the mean-value baseline in all model--benchmark combinations for both CSS and MI.
Across benchmarks, correlations are positive and statistically significant in 15 of 16 model--benchmark combinations for both CSS and MI, showing that PreMaQ generalizes beyond benchmark-specific patterns.
}

\subsection{RQ2: Model Selection Based on Predicted Maintainability-Related Quality}
\label{subsec:rq2}

\begin{figure}[t]
    \centering
    \includegraphics[width=\linewidth]{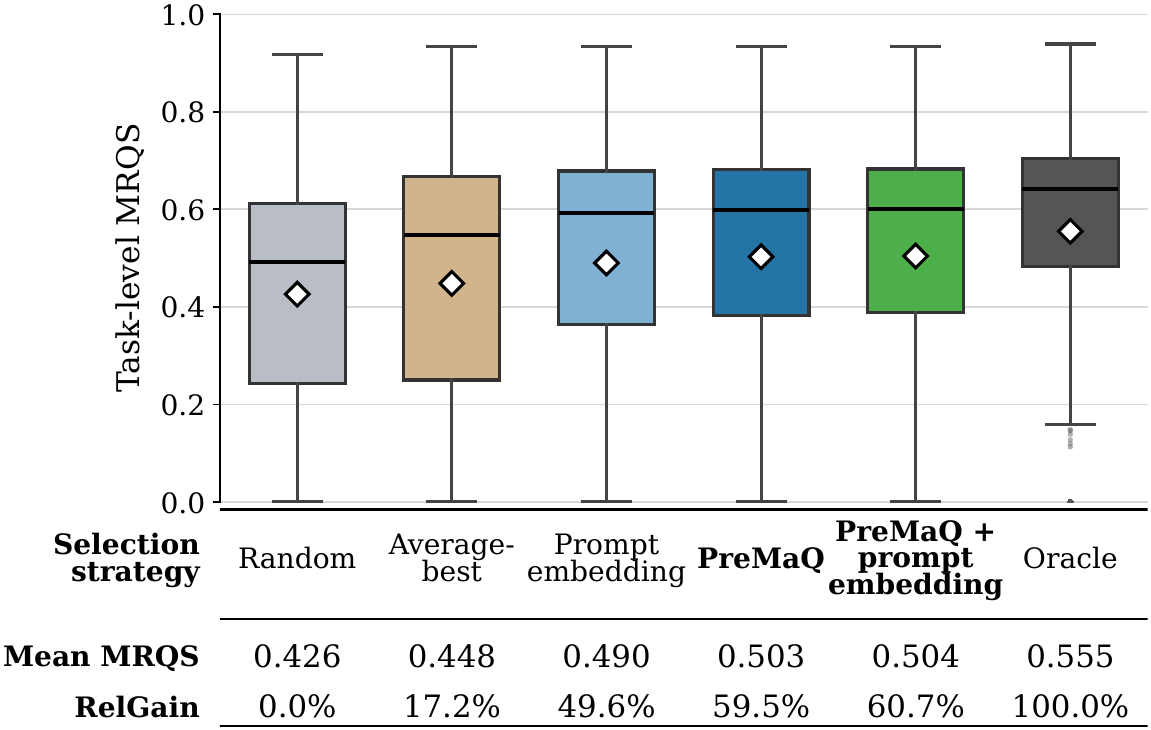}
    \caption{Task-level MRQS distributions achieved by each model-selection strategy on $N{=}1{,}534$ tasks. Diamonds indicate means, and center lines indicate medians. The rows below report mean MRQS and RelGain. Random shows per-task expected MRQS.}
    \label{fig:rq2_boxplot}
\end{figure}

\figref{fig:rq2_boxplot} shows the distributions of MRQS values achieved by each model-selection strategy across tasks and reports the mean MRQS and relative gain for each strategy.

\textbf{PreMaQ-based model selection improves MRQS over the evaluated baselines.}
PreMaQ outperforms Random, Average-best, and prompt embedding, achieving a relative gain of $59.5$\,\% compared with $0.0$\,\%, $17.2$\,\%, and $49.6$\,\%, respectively.
PreMaQ-based model selection improves the maintainability-related quality of generated code without sacrificing functional correctness because every candidate model in the target tasks generates functionally correct code.
The performance of prompt embedding indicates that the task prompt itself can predict maintainability-related quality to some extent.
Notably, combining predictions from PreMaQ and prompt embeddings achieves higher selection performance ($60.7$\,\%) than either method alone.

\begin{figure}[t]
    \centering
    \includegraphics[width=.8\columnwidth]{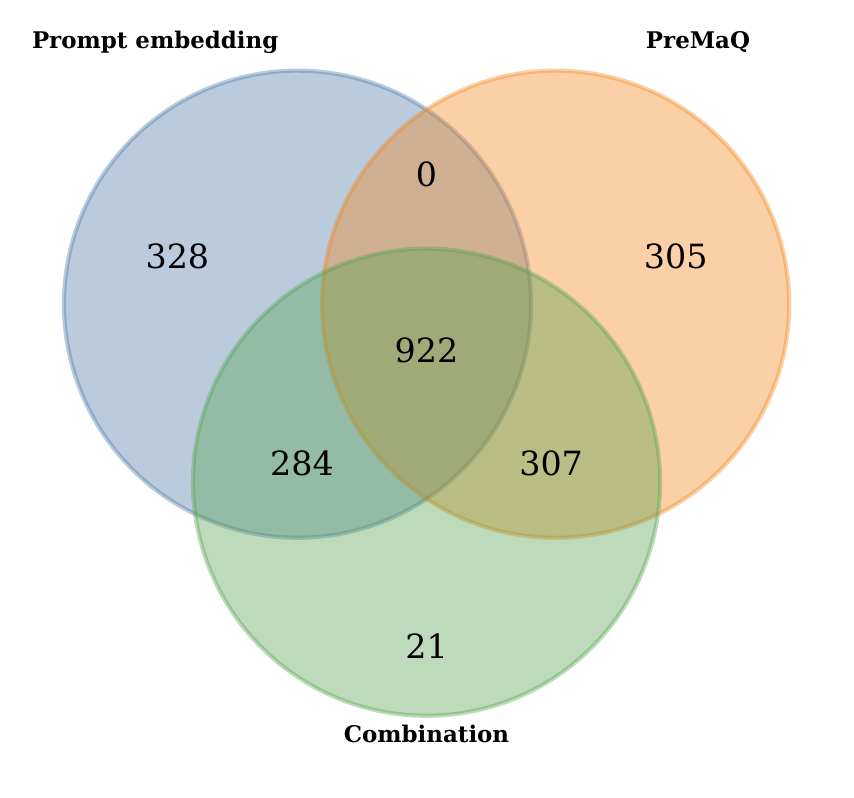}
    \caption{Overlap among the task--model pairs selected by prompt embeddings, PreMaQ, and their combination across 1,534 tasks. Each strategy contributes one pair per task.}
    \label{fig:rq2_venn}
\end{figure}

To examine the complementarity among the three best-performing methods (i.e., PreMaQ, prompt embeddings, and their combination), \figref{fig:rq2_venn} shows the overlap among the task--model pairs selected by these methods.

\textbf{PreMaQ's predictions are complementary to prompt embeddings.}
PreMaQ and prompt embeddings select different models for 612 of the 1,534 tasks ($39.9$\,\%).
Among these tasks, the combination follows the prompt-embedding selection in 284 cases and the PreMaQ selection in 307 cases, while selecting a different model in 21 cases.
Together with the increase in relative gain from $49.6$\,\% to $60.7$\,\%, the highest among all strategies, this result suggests that the two methods provide complementary task-specific signals for model selection.

\summaryblock{Answer to RQ2}{
When multiple LLMs generate functionally correct code for a task, PreMaQ-based selection achieves higher maintainability-related quality than all evaluated baseline strategies without internal representations.
Its relative gain in MRQS is $59.5$\,\%, compared with $17.2$\,\% for Average-best.
Combining PreMaQ and prompt embeddings increases the relative gain to $60.7$\,\%, suggesting that they provide complementary task-specific signals.
}

\section{Maintainability-Aware Model Selection}
\label{sec:discussion}

RQ2 shows that PreMaQ can improve the maintainability-related quality achieved by model selection in a setting with known functionally correct candidates.
This section discusses how maintainability-aware model selection could support developers who use LLMs for code generation.
First, we present a concrete example in which PreMaQ chooses a higher-quality output over another functionally correct output.
We then discuss how maintainability-aware model selection can be integrated with functional-correctness prediction, examine the runtime overhead of using internal representations, and provide practical guidance.

\textbf{\textit{Impact.}} \textbf{A concrete example illustrates an observable quality difference associated with model selection.}
To illustrate this difference, we examine \textit{BigCodeBench/650}, for which three models generate outputs that pass all benchmark tests.
PreMaQ selects Qwen3.5-27B-FP8, whose output follows the standard usage of \texttt{pytz} for timezone localization.
The excerpts below show the lines relevant to this comparison.

\begin{lstlisting}[style=qiskitpython]
dt_naive = parse(date_str)
tz = pytz.timezone(tz_str)
dt_tz = tz.localize(dt_naive)
current_year = dt_tz.year
next_year = current_year + 1
next_new_year_naive = datetime(next_year, 1, 1, 0, 0, 0)
next_new_year_tz = tz.localize(next_new_year_naive)
\end{lstlisting}

We compare this output with the Qwen2.5-Coder-7B output, which passes all tests but has the lowest observed MRQS and contains an unnecessary unreachable branch (Line 5).

\begin{lstlisting}[style=qiskitpython]
given_date = parse(date_str)
given_date = given_date.replace(tzinfo=pytz.timezone(tz_str))
next_year = given_date.year + 1
next_new_year = datetime(next_year, 1, 1, tzinfo=pytz.timezone(tz_str))
if given_date > next_new_year: # unreachable
    next_new_year = datetime(next_year + 1, 1, 1, tzinfo=pytz.timezone(tz_str))
\end{lstlisting}

This example illustrates how PreMaQ can use prefill activations to avoid selecting a model likely to produce less maintainable code before generation.
The complete outputs are provided in our replication package.

\textbf{\textit{Integration.}} \textbf{Maintainability-related quality prediction can be integrated with functional-correctness prediction.}
Prior work~\cite{lugoloobi2026lit,varshney2026arxiv} has used pre-generation predictions of functional correctness to select a model.
These activation-based routing methods provide a direct integration point with our approach.
For example, Lugoloobi et al.~\cite{lugoloobi2026lit} use prefill activations to predict the probability that each candidate model will solve a task and select a model based on the predicted probability and inference cost.
Our predictions of maintainability-related quality can be incorporated into this routing process as an additional selection criterion.
A routing policy could either filter candidates by predicted correctness before considering predicted maintainability-related quality and inference cost, or jointly combine all three criteria according to developer priorities.
Both designs move beyond the RQ2 setting with known functionally correct candidates by using predicted correctness.
This additional prediction of maintainability-related quality does not require another LLM forward pass.
The router already performs a prompt forward pass for correctness prediction, so our prediction requires only capturing the necessary internal representations and running the maintainability-related quality predictors.
The resulting runtime overhead is sufficiently low for practical use, as discussed in the next subsection.

\textbf{\textit{Efficiency.}} \textbf{Internal representations enable efficient maintainability-related quality prediction.}
\begin{table}[t]
\centering
\caption{Runtime of internal-representation-based quality prediction and model selection.}
\label{tab:rq2_selection_overhead}
\begin{tabular}{@{}lr@{}}
\toprule
Component & Time \\
\midrule
Internal representation extraction (4-model avg.) & $145.7$ ms \\
Predictor inference & $0.252$ ms \\
Quality scoring and selection & $1.5$ $\mu$s \\
Predictor training & $14.0$ s \\
\bottomrule
\end{tabular}
\end{table}

As shown in \tabref{tab:rq2_selection_overhead}, in a standalone pipeline after model loading, the dominant per-task runtime cost is the prompt forward pass with internal representation capture, which takes $145.7$ ms on average.
Predictor inference takes only $0.252$ ms, while quality scoring and selection require negligible time.
When integrated with an activation-based router that already derives predictions from prefill activations, the prompt forward pass can be shared, so the additional cost is limited to capturing the required internal representations and running the predictors.
Predictor training is also lightweight in practice, taking only $14.0$ seconds per fold.

\textbf{\textit{Guidance.}} \textbf{PreMaQ is best used alongside prompt embedding when white-box access is available.}
In RQ2, combining predictions from PreMaQ and prompt embeddings achieves the best selection performance, with a relative gain of $60.7$\,\%.
Thus, when internal activations are accessible, PreMaQ should be used as an additional signal alongside prompt-embedding predictions.
For API-only models without access to internal activations, prompt embeddings provide a black-box alternative, which achieves a relative gain of $49.6$\,\% in our evaluation.
Our combination uses a simple average of the two predictions; weighted or learned combinations are left for future work.

\section{Threats to Validity}
\label{sec:threats}

\noindent\textbf{Construct validity.}
This study evaluates maintainability-related code quality using CSS and MI, but these metrics do not cover other quality aspects, such as readability and execution efficiency.
In addition, MRQS, which is used in RQ2, is one way to combine CSS and MI.
Changing how the two metrics are combined may change the selection results.

\noindent\textbf{Internal validity.}
The code generated by an LLM depends on the prompt format.
To reduce quality degradation caused by arbitrary prompt formats, we followed the official prompt format provided by each benchmark.
The prediction results also depend on the configuration of the MLP predictor, and using different settings may change prediction performance.

\noindent\textbf{External validity.}
The results of this study are based on four local models and four Python code generation benchmarks.
To improve generalizability, we included both general-purpose and code-specialized models, evaluated benchmarks with different levels of complexity, and further conducted cross-benchmark evaluation.
However, it remains untested whether the results generalize to larger models, different model families, or programming languages other than Python.
The prompt embedding baseline uses only all-MiniLM-L6-v2, and using a different embedding model may change the comparison results.

RQ2 uses a setting with known functionally correct candidates.
Section~\ref{sec:discussion} outlines integration with pre-generation functional-correctness prediction, but we do not evaluate it end to end.
The quality improvement reported in RQ2 may vary depending on prediction accuracy and the routing policy.

\section{Conclusion}
\label{sec:conclusion}

This study proposed PreMaQ, which predicts the maintainability-related quality of LLM-generated code from prefill activations, and evaluated its use in model selection.
Using four open-weight LLMs and four Python code generation benchmarks, we found that predictions of CSS and MI are correlated with their observed values in all within-benchmark model--benchmark combinations.
Prediction also generalizes across benchmarks for most combinations, more consistently for MI than for CSS.
In the RQ2 setting with known functionally correct candidates, PreMaQ-based model selection outperforms the evaluated strategies without internal representations, while combining predictions from PreMaQ and prompt embeddings achieves the best selection performance.

These findings suggest that developers could consider predicted maintainability-related quality alongside predicted functional correctness and inference cost when selecting a model before code generation.
For researchers, prior work and our findings motivate future investigation into which additional generated-code properties can be predicted from internal representations.
Such investigation could reveal the broader potential of internal representations as pre-generation signals for software engineering.

\section{Data Availability}
All data and code used in this study are available at the following repository: https://github.com/posl/PreMaQ-replication.git.
\section*{Acknowledgment}
We gratefully acknowledge the financial support of: (1) JSPS for the KAKENHI grants (JP24K02921, JP25K03100, JP25K22845, JP26H02500, JP26K25555); (2) Japan Science and Technology Agency (JST) as part of Adopting Sustainable Partnerships for Innovative Research Ecosystem (ASPIRE), Grant Number JPMJAP2415, and (3) the Inamori Research Institute for Science for supporting Yasutaka Kamei via the InaRIS Fellowship.

\bibliographystyle{IEEEtran}
\bibliography{references}

\end{document}